\documentclass[sigconf]{acmart}

\usepackage{graphicx}
\usepackage{balance}  
\usepackage{makecell}
\usepackage[font=bf]{caption}
\usepackage[font=bf]{subcaption}
\usepackage{amsmath}
\usepackage{etoolbox}
\usepackage{url}
\usepackage{amssymb}
\usepackage{diagbox}
\usepackage{multirow}
\usepackage{hhline}
\usepackage{listings}
\usepackage{color}
\usepackage{enumitem}
\usepackage{float}
\usepackage{comment}
\usepackage{etoolbox}

\AtBeginDocument{%
  \providecommand\BibTeX{{%
    \normalfont B\kern-0.5em{\scshape i\kern-0.25em b}\kern-0.8em\TeX}}}

\begin{document}

\newcommand{\nvm}{PMem}
\newcommand{\ssd}{NVMe SSD}
\newcommand{\sssd}{SSD}
\newcommand{\sector}{sector}
\newcommand{\fsdaxfull}{FileSystem-DAX}
\newcommand{\devdaxfull}{Device-DAX}
\newcommand{\fsdax}{FS-DAX}
\newcommand{\devdax}{Dev-DAX}

\newcommand{\puredax}{\nvm\ FS-DAX mode direct I/O}
\newcommand{\daxfio}{\nvm\ FS-DAX mode File I/O}
\newcommand{\stfio}{\nvm\ sector mode File I/O}
\newcommand{\ssdfio}{\ssd\ File I/O}
\newcommand{\daxio}{DAX}
\newcommand{\fileio}{File I/O}
\newcommand{\appinterl}{AppDirect Interleaved}
\newcommand{\appnoninterl}{AppDirect NonInterleaved}
\newcommand{\appmode}{AppDirect Mode}
\newcommand{\memmode}{Memory Mode}
\newcommand{\pmemmem}{Mem}

\newcommand{\fiveminrule}{The Five Minute Rule}

\newcommand{\REMARK}[2]{{\bf [*[#1:~#2]*]}}
\newcommand{\cmt}[1]{\REMARK{comment}{#1}}
\newcommand{\mmdram}{MM-DRAM}
\newcommand{\mmcache}{MM-Cached}
\newcommand{\tpcc}{TPC-C}
\newcommand{\tpch}{TPC-H}
\newcommand{\sqlserver}{Sql Server}

\newcounter{foo}
\newenvironment{foo}[1][]{\refstepcounter{foo}\par
   \textbf{Observation ~\thefoo. #1} \rmfamily}{}

\newcounter{factor}
\newenvironment{factor}[1][]{\refstepcounter{factor}\par\medskip
   \textbf{Factor ~\thefactor. #1} \rmfamily}{\medskip}
   
\newcounter{challenge}
\newenvironment{challenge}[1][]{\refstepcounter{challenge}\par\medskip
   \textbf{Challenge ~\thechallenge. #1} \rmfamily}{\medskip}
\newcommand{\yinjun}[1]{\textcolor{red}{[yinjun: #1]}}
\newcommand{\kp}[1]{\textcolor{blue}{[KP: #1]}}
\newcommand{\rs}[1]{\textcolor{brown}{[RS: #1]}}
\newcommand{\bpk}[1]{\textcolor{olive}{[BPK: #1]}} 
\newcommand{\jae}[1]{\textcolor{purple}{[JAE: #1]}}

\newcommand{\code}[1]{\texttt{#1}}

\title{Lessons learned from the early performance evaluation of Intel Optane DC Persistent Memory in DBMS}

\author{Yinjun Wu{ $^{1}$}, Kwanghyun Park{ $^{2}$}, Rathijit Sen{ $^{2}$}, Brian Kroth{ $^{2}$}, Jaeyoung Do{ $^{3}$}}
 \affiliation{
 $^{1}$University of Pennsylvania,
$^{2}$Microsoft Gray Systems Lab,
$^{3}$Microsoft Research\\
\texttt{wuyinjun@seas.upenn.edu}\\
\texttt{\{<name>.<surname>\}@microsoft.com}\\
\texttt{jaedo@microsoft.com}\\
}
\begin{abstract}

Non-volatile memory (NVM) is an emerging technology, which has the persistence characteristics of large capacity storage devices (e.g., HDDs and SSDs), while providing the low access latency and byte-addressablity of traditional DRAM memory.
This unique combination of features open up several new design considerations when building database management systems (DBMSs), such as replacing DRAM (as the main working space memory) or block devices (as the persistent storage), or complementing both at the same time for several DBMS components (such as access methods, storage engine, buffer management, logging/recovery, etc).

However, interacting with NVM requires changes to application software to best use the device (e.g. \code{mmap} and \code{clflush} of small cachelines instead of \code{write} and \code{fsync} of large page buffers).
Before introducing (potentially major) code changes to the DBMS for NVM, developers need a clear understanding of NVM performance in various conditions to help make better design choices.

In this paper, we provide extensive performance evaluations conducted with a recently released NVM device, Intel Optane DC Persistent Memory (PMem), under different configurations with several micro-benchmark tools. Further, we evaluate OLTP and OLAP database workloads (i.e., \tpcc\ and \tpch) with Microsoft SQL Server 2019 when using the NVM device as an in-memory buffer pool or persistent storage.
From the lessons learned we share some recommendations for future DBMS design with \nvm
, e.g. simple hardware or software changes are not enough for the best use of \nvm\ in DBMSs. 

\end{abstract}

\maketitle

\definecolor{dkgreen}{rgb}{0,0.6,0}
\definecolor{gray}{rgb}{0.5,0.5,0.5}
\definecolor{mauve}{rgb}{0.58,0,0.82}

\lstset{frame=tb,
  language=C,
  aboveskip=3mm,
  belowskip=3mm,
  showstringspaces=false,
  columns=flexible,
  basicstyle={\small\ttfamily},
  numbers=none,
  numberstyle=\tiny\color{gray},
  keywordstyle=\color{blue},
  commentstyle=\color{dkgreen},
  stringstyle=\color{mauve},
  breaklines=true,
  breakatwhitespace=true,
  tabsize=3
}

\section{Introduction}
Non-volatile memory (NVM) is an emerging technology which forces the database community to revisit various DBMS internals (e.g., access methods, storage engine, buffer management, logging, recovery, etc.) \cite{arulraj2019non} because of its unique device characteristics. For example, one recently released NVM device, Intel Optane DC persistent memory (PMem for short hereafter), provides large capacity and persistence like traditional block storage devices (e.g., HDDs and SSDs), as well as low latency and byte-addressability like DRAM memory. 


Over the last few years, there have been some efforts on measuring the performance of this new device \cite{izraelevitz2019basic, hpe2019pmem, lenovo2019pmem, lenovo2019pmem2}.
However, some important \nvm\ device measurements, which is critical to the performance of processing database queries, are missing (e.g., degree of parallelism, I/O request size, etc.).
Further, some important database design questions that need to be considered when developing DBMSs for NVM have not been thoroughly answered. For example, due to its unique characteristics, \nvm\ can be possibly used in DBMSs for overcoming the limited DRAM buffer pool size or for improving the performance of persistent storage or both. Such system design choices still remain largely open.

The focus of this paper is to explore PMem's characteristics, mainly in the DBMS perspective, under several \nvm\ configurations with  
micro-benchmark tools.
The observations and analysis we gain in this paper help explain more traditional database workload performance results (e.g., impact of writes on OLTP and OLAP workloads) and inform our suggestions for optimal DBMS internal parameters for query processing on PMem.
For example, our results show that the small I/O request sizes can best take advantage of \nvm's relatively high performance compared to other traditional storage devices (i.e., \sssd s). Additionally, we observe that \nvm\ devices do not exhibit the same concurrent request scaling rules as SSDs.

In addition, we evaluate OLTP and OLAP workloads (i.e., TPC-C and TPC-H) with different \nvm\ device configurations to explore the potential database design choices when integrating \nvm\ in DBMSs.
Our OLTP and OLAP 
evaluations reveal some important design considerations when building DBMSs with \nvm:\ 
(1) Replacing a traditional DRAM based buffer pool with \nvm\
configured in Memory Mode
(e.g., for its increased capacity for working set memory and ability to use directly without software changes) is a suboptimal solution because of its extreme performance asymmetry between reads and writes.
This design choice hurts not only the performance of write intensive OLTP workloads, but also that of read intensive OLAP workloads because all intermediate query results are also written to PMem;
(2) extending a DRAM buffer pool with \nvm\ requires a PMem enlightened database page to buffer placement policy.
For example, our results show that the performance of OLAP workloads can drop significantly if we do not ensure hot pages land on DRAM, despite the high read performance of PMem.

Our device micro-benchmark results and database workload evaluations point out that the solutions introduced so far only provided initial implementations, but there are huge opportunities for additional research to properly adopt this new technology for DBMS internals.
  
In summary, the contributions of our paper are as follows:
\begin{itemize}
    \item We share some important \nvm\ device characteristics and introduce crucial factors (e.g., degree of parallelism, I/O request size, and impact of frequent writes) for database query processing with \nvm.
    \item We evaluate OLTP and OLAP workloads with different \nvm\ configurations (i.e., replacing DRAM buffer pool, extending DRAM buffer pool, and replacing storage devices with \nvm) in a production grade DBMS, and
    \item Based on our findings, we make design choice recommendations for buffer management and storage engines for DBMSs.
\end{itemize}

In the remainder of the paper, we present related work and different possible \nvm\ configurations in Section \ref{sec: related_work} and Section \ref{sec: pmem_intro} followed by the \nvm\ device performance and database workload evaluations with \nvm\ in Section \ref{sec: device_character} and \ref{sec: query_processing} respectively.
Some future research directions are discussed in Section \ref{sec: conclusion}.

\section{Related work}\label{sec: related_work}

After the recent release of \nvm, initial efforts on performance characterization of the device have been started \cite{izraelevitz2019basic, yang2019empirical, hpe2019pmem, lenovo2019pmem, weiland2019early, peng2019system},
which also motivates the performance evaluations of \nvm\ in different applications, e.g. B-tree performance evaluations \cite{lersch2019evaluating}, scientific benchmark evaluations \cite{mironov2019performance}, power usage evaluations \cite{peng2019system}, hybrid memory system evaluations \cite{imamura2020analysis}, integer compression schemes \cite{zarubin2019integer}  and some initial database workload evaluations on \nvm\ \cite{izraelevitz2019basic}. There are also some other initial studies on how to close the latency gap between DRAM and \nvm\ in latency-sensitive operations \cite{psaropoulos2019bridging}, how to provide efficient I/O primitives with \nvm\ \cite{van2019persistent}, how to design better page replacement policy with \nvm\ \cite{lersch2019persistent} and how to efficiently provide replication mechanisms with \nvm\ \cite{zarubin2019efficient}.
However, to our knowledge, no detailed database workload evaluations and DBMS design recommendations on the newly released \nvm\ device exist yet.

Prior to the release of \nvm, a series of research works have been motivated by the predicated use of future NVM devices.
These efforts focused on how to utilize NVM in database systems for improved performance, recovery, or other lines compared to other storage devices (see \cite{arulraj2016write, chen2015persistent, viglas2014write, blelloch2015sorting, van2018managing} and the references herein).
However, due to the lack of real NVM devices when those works were published, they were based on NVM emulated in DRAM, typically with an expectation of 
near DRAM bandwidths, which we now know to be inaccurate assumptions.
Hence, the feasibility of those ideas on the real NVM device is still unknown.




\section{\nvm\ configurations}\label{sec: pmem_intro}
\label{sec:pmemconf}

\begin{figure}[t]
     \centering
     \includegraphics[width=0.5\textwidth]{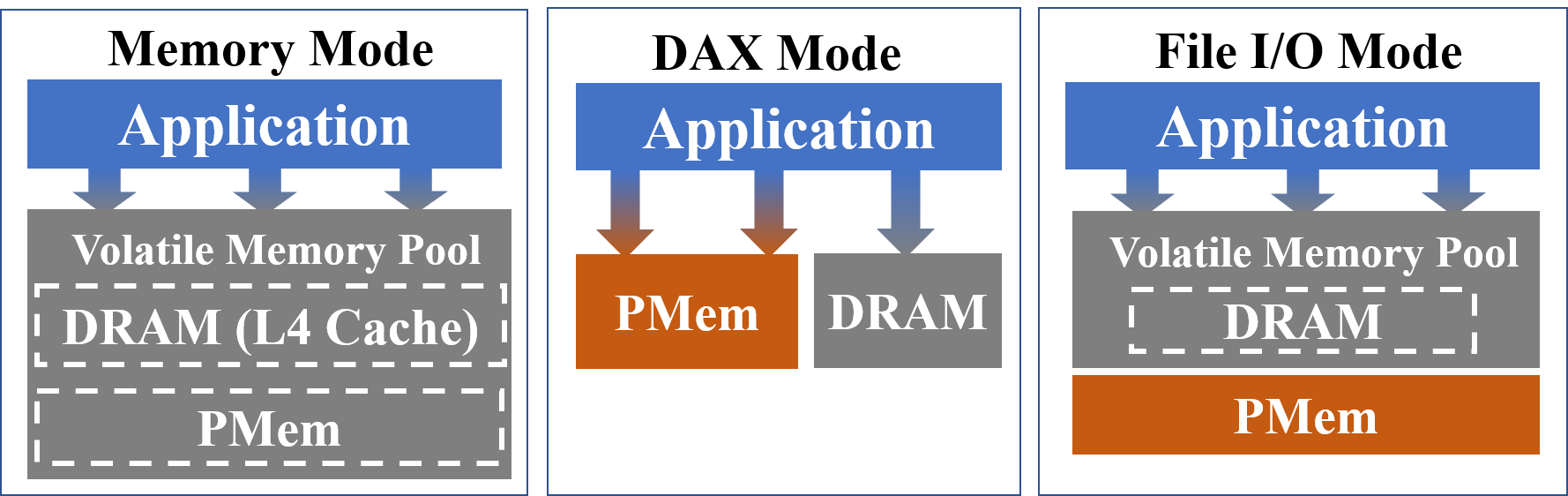}
     \caption{Different configurations of \nvm. PMem with the orange color (in the \nvm\ and \fileio\ modes) indicates the use of PMem as persistent storage while the gray color (in the Memory Mode) means its usage as volatile memory.}
     \label{fig:DAX_mem_comp}
\end{figure}

We use three different modes (shown in Figure \ref{fig:DAX_mem_comp}) of accessing PMem, each of which has its own set of trade-offs and must be chosen in the system's firmware configuration.

First, in \emph{Memory Mode} PMem is used as the main memory of a system that leverages DRAM as a high-speed L4 cache, managed by the hardware's memory controller.
This mode can be used as a way to increase the capacity of the working memory for a system, and to use PMem with existing applications without modification.
However, in this mode, PMem is actually still volatile.
That is to say, all data written to PMem in Memory Mode is lost upon a system restart (whether the cause of the restart is intentional or not).
Thus, in this mode we are not able to take advantage of a key feature of the device.
Moreover, the cache (re)placement policy is subject to NUMA and other effects and cannot be managed by software.

In both of the other access mode options, \emph{Direct Access (DAX)} and \emph{File I/O}, PMem guarantees persistence of the data it stores.

In \emph{DAX mode}, applications use PMem via memory semantics (i.e., \code{load} and \code{store} instructions), after an initial interaction with the OS kernel through the \code{mmap} syscall to setup a virtual address space mapping, thus allowing the direct CPU data access to PMem's address space without further kernel intervention.
However, this may involve significant changes to application code to handle things like torn writes due to CPU cacheline flushing semantics that most applications don't generally need to be concerned with.
Instead, application developers have to carefully inject \code{clflush}, \code{mfence}, and other low level instructions or make use of libraries (e.g. Intel's PMDK \cite{pmdk}), which are known to introduce other inefficiencies \cite{kadekodi2019splitfs}, to do it for them.

On the other hand, in \emph{File I/O mode}, PMem is accessed by applications using standard file system APIs (e.g., \code{read}, \code{write}, etc.).
This has higher latency than DAX mode when the number of data accesses within a region are frequent since each operation must pass through the entire I/O software stack of the OS.
Additionally, block I/O typically does not provide byte-addressable semantics.

Table \ref{table:config} provides a summary of the PMem configurations.

\begin{table}[h]
\begin{center}
\begin{tabular}{|c|c|c|c|c|c|}\hline
\multicolumn{2}{|c|}{\backslashbox{feature \strut}{\strut configuration}} & \makecell{Memory \\ Mode} & \makecell{\daxio\ \\ Mode} & \makecell{\fileio\ \\ Mode} \\\hline
\multirow{2}{*}{\makecell{disk like \\ features}} & Persistence & & \checkmark & \checkmark \\\hhline{~----}
 & \makecell{Large  capacity} &\checkmark & \checkmark & \checkmark \\\hline
\multirow{2}{*}{\makecell{memory like\\ features}} & \makecell{Byte \\ addressability} &\checkmark & \checkmark &  \\\hhline{~----}
 & \makecell{CPU \\ direct access} & \checkmark & \checkmark & \\\hline
\end{tabular}
\caption{A summary of different \nvm\ configurations.}
\label{table:config}
\end{center}
\end{table}
\section{\nvm\ micro-Benchmark Analysis}\label{sec: device_character}

This section presents a comprehensive performance analysis of PMem device characteristics when used as persistent storage (i.e., the DAX or File I/O modes) or as main memory (i.e., Memory Mode), as described in Section~\ref{sec:pmemconf}.

\subsection{System configuration}\label{sec: sys_config}
We conduct all experiments on a dual-socket server with Intel Xeon Platinum 8260L CPUs.
Each socket has 24 physical cores, each of which has its own 32KB L1 and 1MB L2 caches, and a shared 35.75MB L3 cache.
Each socket is also populated with 96GB of DRAM (six 16GB Micron DDR4), and 1
TB PMem (two interleaved 512GB Intel Optane DC Persistent Memory modules installed in memory DIMM slots).

Note that all experiments in this section are performed on the two interleaved PMem modules in one socket accessed by the local CPUs (i.e., no remote NUMA accesses).
To compare performance when PMem is used as persistent storage, we use one NVMe 4TB Intel DC P4510 Series SSD (called \sssd\ for short hereafter).
Finally, we run Ubuntu 18.10 on the server where hyper-threading is enabled, and the CPU power governor is configured to performance mode, forcing the CPU to use the highest possible (turbo) clock frequency.

\subsection{\nvm\ as persistent storage}\label{sec: micro_bench_storage}


We first investigate the performance characteristics of \nvm\ when it is used in the \daxio\ or \fileio\ modes, comparing with \sssd, and study their implications when designing DBMSs for \nvm.
For this evaluation, we use Flexible I/O tester (\code{fio}) \cite{FIO315} to issue synchronized I/O requests, varying several parameters, including the I/O request size\footnote{In the context of DAX mode we mean \code{load}/\code{store} instruction operations when we refer to "I/O requests", whereas in block mode we mean block read/write operations.}, access patterns (random/sequential and read/write), and the number of I/O threads, seeking to answer the following questions:

\begin{enumerate} [start=1,label={(\bfseries Q\arabic*)}]
\item \label{Q: IO_app} How will the different I/O sizes affect the performance of \nvm\ in the \daxio\ and \fileio\ modes?
\item \label{Q: Thread_app} How is \nvm's performance affected as the number of I/O threads increases?
\item \label{Q: rw_app} What is the performance difference between read and write requests in \nvm?
\end{enumerate}

\begin{figure}[t]
\centering
\begin{subfigure}{0.5\textwidth}
\centering
        \includegraphics[width=0.85\textwidth]{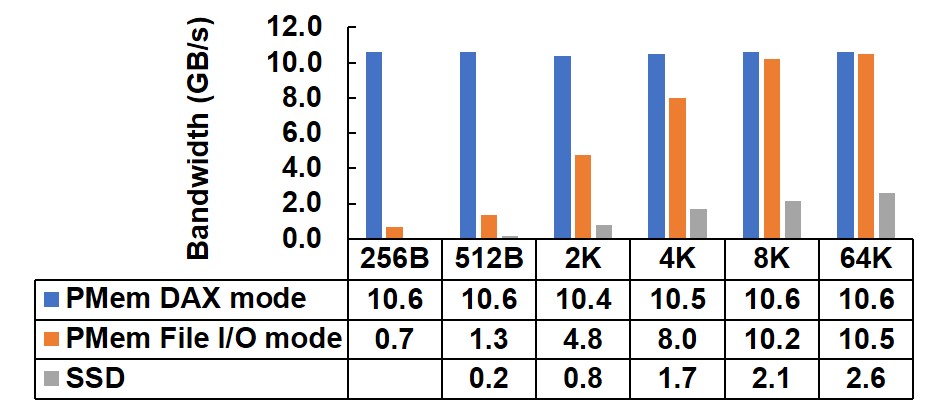}
    \caption{Read}    
    \label{fig:sr_bw}
    \end{subfigure}
    \begin{subfigure}{0.5\textwidth}
\centering
        \includegraphics[width=0.85\textwidth]{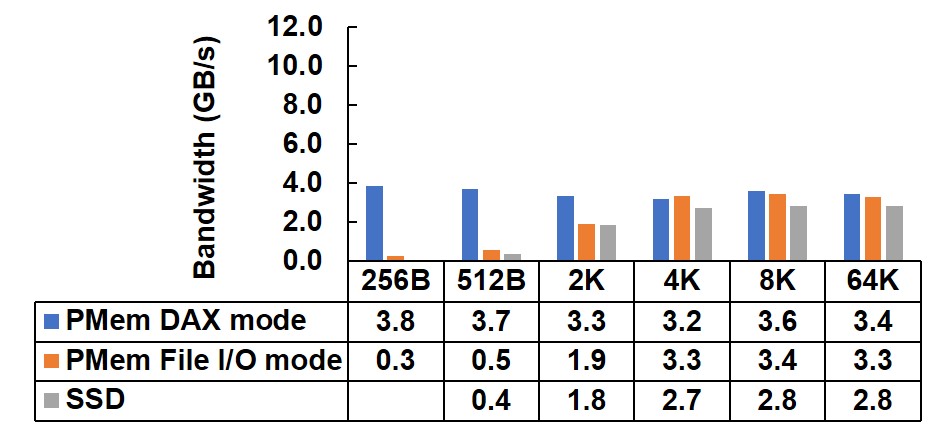}
    \caption{Write}    
    \label{fig:sw_bw}
    \end{subfigure}
    \caption{Sequential-access bandwidth results. Note that the minimum I/O size for \sssd\ is limited to its device sector size (i.e., 512B) while there is no such limitation for \nvm\ in \daxio\ and \fileio\ modes.}
    \label{fig:bw}
\end{figure}

\begin{figure}[t]
\centering
    \begin{subfigure}{0.5\textwidth}
    \centering
     \includegraphics[width=0.7\textwidth, height=0.35\textwidth]{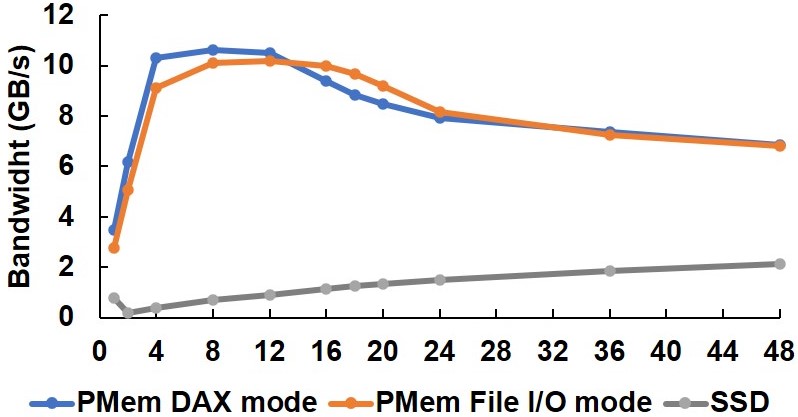}
    \caption{Read}    
    \label{fig:r_bw_thread}
    \end{subfigure}
    \begin{subfigure}{0.5\textwidth}
    \centering
     \includegraphics[width=0.7\textwidth, height=0.35\textwidth]{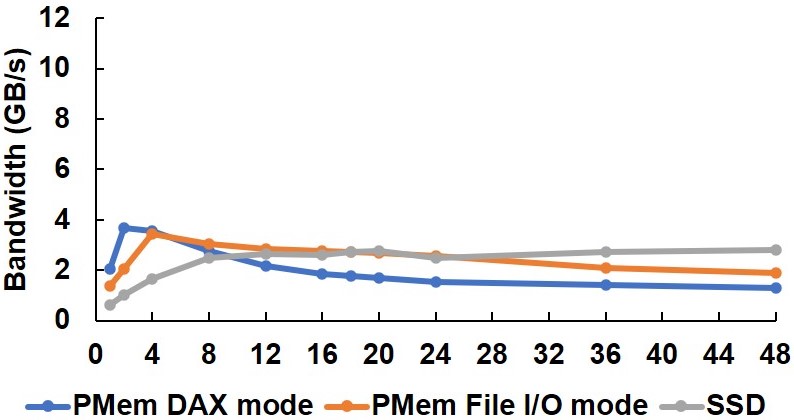}
    \caption{Write}    
    \label{fig:w_bw_thread}
    \end{subfigure}
    \caption{Sequential-access bandwidth results as varying the number of I/O threads, each of which issues 8KB reads and writes, respectively.}
    \label{fig:bw_thread}
\end{figure}

While previous work \cite{izraelevitz2019basic} has already addressed portions of the above questions (e.g., \cite{izraelevitz2019basic} focuses on single-thread experiments while we vary the number of I/O threads and I/O sizes simultaneously), our main contribution in this analysis is to consider how PMem's performance characteristics affect when redesigning various DBMS components for \nvm.
Note that due to space limits, we present bandwidth results measured with only sequential \mbox{read/write} workloads (Figure \ref{fig:bw}--\ref{fig:bw_thread}). We also measured the latency results, which, however are less relevant to the database workload experiments and thus not included here.

\begin{foo}\label{observ: small_IO}
\textit{Figure \ref{fig:bw} shows the peak bandwidth of read and write workloads with varied I/O request sizes. When the I/O size is small (i.e., $<4$KB), \nvm\ \daxio\ shows significant performance gains compared to \nvm\ \fileio\ and \sssd\ (e.g., with 512B I/O size, \nvm\ \daxio\ achieves about 8x and 50x higher throughputs than \nvm\ \fileio\ and \sssd, respectively). However, with bigger I/O sizes, there is no distinct performance advantage of \nvm\ \daxio\ compared to \nvm\ \fileio.}
\end{foo}



\textbf{Analysis} Unlike \nvm\ \daxio\, of which the internal device access size is 256B with 64B cacheline operations from the byte level CPU \code{load}/\code{store} instructions perspective, \nvm\ \fileio\ and \sssd\ access data in the sector size granularity exposed to the OS, which is typically configured to 4KB or 512B.
Therefore, \nvm\ \fileio\ and \sssd\ waste I/O bandwidth if the required data size is smaller than the block size.
Further \nvm\ \fileio\ and \sssd\ always need to go through the I/O stack to fetch data, adding extra system call overheads.
On the other hand, this overhead can be easily amortized when issuing large I/Os, resulting in no noticeable advantage in \nvm\ \daxio\ over others.

\textbf{Recommendations} 
The performance improvement achieved by using small I/O sizes in \nvm\ \daxio\ provides an interesting opportunity to speed up the query processing in DBMSs with small page sizes. As an example, let's consider using a 512B page instead of 8KB for a B-tree node, and further assume that each record has a size of 64B. Then we can fit 8 and 128 records into 512B and 8KB nodes, respectively, and the depth of the B-tree index can be calculated as $log_{8}(n)$ for the 512B node, and $log_{128}(n)$ for the 8KB node, where $n$ is the total number of records in the database. Thus, the depth difference between two node sizes would be $log_{8}(n)/log_{128}(n)\approx2$. Given the latency of accessing 512B data is about five times faster than 8KB (i.e., on a single-thread measurement, we observed that the access latency in \nvm\ \daxio\ is reduced from 2.6 $\mu$s to 0.5 $\mu$s), we can expect in total about 2.5x faster index lookup time of traversing from the root to a leaf node (i.e., 2x deeper depth, but \textasciitilde5x faster page access latency) with the 512B node.

\begin{foo}\label{ob: high_parall}
\textit{Figure \ref{fig:bw_thread} shows how read/write bandwidth changes with the number of I/O threads using a fixed-size I/O size (i.e., 8KB).
As can be seen in the figure, the read/write bandwidth of \nvm\ gradually drops after a certain number of threads (i.e., 12 and 4 threads in reads and writes, respectively).
Thus, although \nvm\ provides better overall throughput than \sssd, it exhibits poor concurrent request scaling compared with it.}
\end{foo}

\textbf{Analysis}: With more threads, the per-thread sequential access pattern becomes in aggregate more random from a device perspective, thus causing more congestion at the internal \nvm\ device controller.
(A similar observation can be found in \cite{izraelevitz2019basic}).
The results shown in Figure~\ref{fig:bw_thread} indicate that, unlike \sssd, a large number of outstanding I/Os are not needed to fully exploit the parallelism of \nvm\ for both modes.


\textbf{Recommendations} The relatively poor scalability of \nvm\ compared to \sssd\ recommends us to limit the maximum degree of parallelism for \nvm\ I/O within a database engine.

\begin{foo}
\textit{The performance of reads and writes of the \daxio\ and \fileio\ are asymmetric (peak read bandwidth vs. peak write bandwidth for both \nvm\ \daxio\ and \fileio: \textasciitilde10.6GB/s vs. \textasciitilde3.5GB/s). In contrast, \sssd\ provides more balanced performance between reads and writes (\textasciitilde2.7GB/s vs. \textasciitilde2.8GB/s). As mentioned earlier, this performance pattern is also observed by \cite{izraelevitz2019basic}.
}
\end{foo}

\textbf{Recommendations} The read-write asymmetry of \nvm\ implies the necessity of avoiding writes as much as possible for \nvm.

\subsection{\nvm\ as Memory}\label{sec: perf_character_mem}

In this subsection, we present the performance characteristics of \nvm\ in memory mode by using Memory Latency checker (MLC hereafter) \cite{mlc}, which is compared with the performance of DRAM and the \nvm\ in \daxio\ mode.

Also as observed in \cite{weiland2019early}, performance measurements in \memmode\ require careful attention to data sizes.
Too small, and they will fit entirely in the DRAM L4 cache, and thus won't measure \nvm\ behavior at all.
To address this we initialized two different size of data (100GB and 500GB) in \nvm\, each of which is larger than the 96GB DRAM we have allocated to a single socket.\footnote{Recall that measurements are restricted to a single socket.}
Note that during the write performance measurements on 500GB dataset in \nvm\ \memmode, MLC crashed due to some internal issues.
So, we present the write performance only for 100GB dataset.
In this subsection, we wish to address the following questions:


\begin{enumerate} [start=4,label={(\bfseries Q\arabic*)}]
\item \label{Q: thread_mem} How does the number of threads influence \nvm\ \memmode\ performance? 
\item \label{Q: workset_mem} How does the dataset size influence \nvm\ \memmode\ performance?
\item \label{Q: comparison_mem} What is the relative performance of \nvm\ \memmode\ compared to that of \nvm\ \daxio\ and DRAM?
\end{enumerate}

Note that \cite{lenovo2019pmem2} and \cite{weiland2019early} have provided initial evaluations on \nvm\ \memmode\ with different dataset sizes under different system configurations, addressing \ref{Q: workset_mem} and also partially \ref{Q: comparison_mem}.
But the influence of the thread count against the performance is still missing.

Due to the space limit, only the sequential read and write bandwidth are shown in Figure~\ref{fig:mlc_res}, from which we note the following:


\begin{figure}[t]
    \centering
    \begin{subfigure}{0.5\textwidth}
    \centering
     \includegraphics[width=0.80\textwidth]{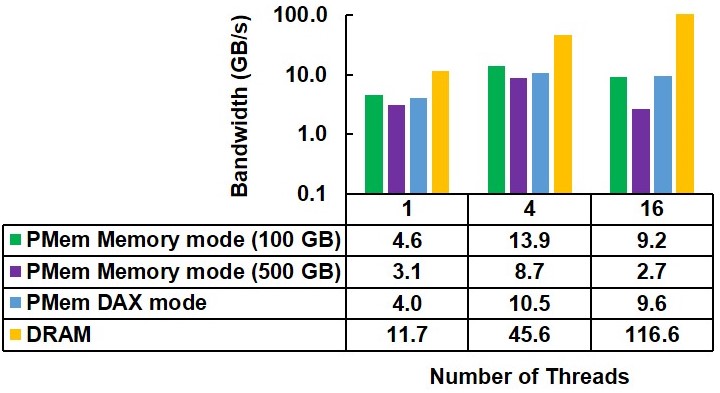}
    \caption{Read}    
    \label{fig:mlc_res_r}
    \end{subfigure}
    \begin{subfigure}{0.5\textwidth}
    \centering
     \includegraphics[width=0.80\textwidth]{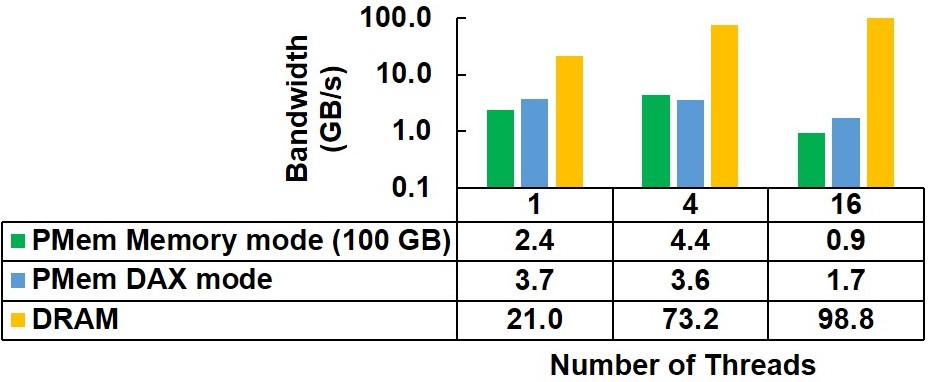}
    \caption{Write}    
    \label{fig:mlc_res_w}
    \end{subfigure}
    \caption{Sequential read/write bandwidth results when \nvm\ is used as memory.}
    \label{fig:mlc_res}
\end{figure}

\begin{foo}
\textit{As the number of threads increases, the \nvm\ \memmode\ bandwidth increases first before it becomes saturated and drops.}
\end{foo}

\textbf{Analysis}: This performance pattern also appears in \nvm\ \daxio\ and \nvm\ \fileio, which can thus be explained in the same way as Observation \ref{ob: high_parall} in Section \ref{sec: micro_bench_storage}.
Namely, concurrent requests appear as random I/O resulting in increased contention to the device controller.

\textbf{Recommendations} Similar to the discussion in Observation~\ref{ob: high_parall}, to use \nvm\ as the buffer pool in a DBMS via \memmode, it is advisable to limit the number of concurrent threads accessing \nvm.
However, in this context, especially when the working set exceeds the size of DRAM, it means limiting concurrency inside the \emph{entire buffer pool} as well, since we lack other software control over which pages are in the L4 DRAM cache vs \nvm, thus effectively limiting the amount of concurrency on the entire DBMS.

\begin{foo}\label{ob: mem_data_size}
\textit{\nvm\ \memmode\ read bandwidth drops as the size of data increases (e.g., the peak bandwidth drops from \textasciitilde13.9GB/s (100GB) to \textasciitilde8.7GB/s (500GB) with four threads)} 
\end{foo}

\textbf{Analysis}: Similar performance pattern is also observed in \cite{weiland2019early}:
as the dataset size increases in \nvm\ \memmode, a larger proportion of requests miss in the L4 DRAM cache, and have to be serviced by \nvm.
As noted previously in Observation~\ref{ob: high_parall}, this will result in a larger number of concurrent accesses to DRAM which may appear more random and result in more contention.

\begin{foo}\label{ob: mem_comparison}
\textit{Under write workloads with smaller data size (100GB), the \nvm\ \memmode\ performs slightly better than \nvm\ \daxio\ while with larger data size or under highly parallel write workloads with smaller data size,
\nvm\ \memmode\ performs worse than \nvm\ \daxio. We can also observe a huge gap between the DRAM performance and \nvm\ performance}
\end{foo}


\textbf{Analysis}: As explained above, in \nvm\ \memmode, if the data size in \nvm\ is larger, DRAM page misses may happen, thus slowing down the data access speed, which becomes more severe for writes in \nvm\ \memmode\ (see the worse write performance of \nvm\ \memmode\ than that of \nvm\ \daxio\ with 16 threads in Figure \ref{fig:mlc_res_w}). 
%
In contrast, 
DRAM misses do not happen for \nvm\ \daxio\ due to the CPU direct access. 
Also as expected, the DRAM performance is much better than \nvm. 

\textbf{Recommendations} Through the observation \ref{ob: mem_data_size} and observation \ref{ob: mem_comparison}, we know that with data size larger than DRAM, especially under write-intensive workloads, \nvm\ \daxio\ performs better than \nvm\ \memmode.
As we will see in Section \ref{sec: query_processing}, under OLAP workloads where many writes for intermediate results are issued, \nvm\ \memmode\ still perform worse than \nvm\ \daxio, and even worse than \sssd.

\section{Database workload evaluations}\label{sec: query_processing}

We will now present our performance evaluation of typical OLTP and OLAP workloads, specifically, {\tpcc} and {\tpch} workloads, with different \nvm\ configurations in Microsoft SQL Server 2019.


\subsection{Experimental design}



As mentioned earlier, we can directly use \nvm\ in SQL Server 2019
either as the persistent storage or 
in conjunction with the buffer pool.
For the former use case, \fileio\ is currently still issued against {\nvm}.
For the latter use case, the DRAM buffer pool in SQL Server 2019 can be extended with Hybrid buffer pool support \cite{hybrid_pmem}, whereby the warm pages cached in \nvm\ are accessed with \daxio\ without being cached in DRAM buffer pool.

In the following, we continue to use {\em \nvm\ \daxio} and {\em \nvm\ \fileio} to denote the use of \nvm\ as the persistent storage, with the Hybrid buffer pool enabled and disabled respectively. 
We compare these with the traditional DBMS configuration: \sssd\ as the persistent storage and DRAM as the buffer pool (denoted as {\em \sssd}). 
When \nvm\ is used in {\memmode} as the buffer pool in SQL Server, DRAM behaves like an L4 cache and \sssd\ is the persistent storage, which we denote by {\em \sssd\ + \nvm\ \memmode}. 
For this case, we denote the traditional DBMS configuration as {\em \sssd\ + DRAM} to highlight the difference in the two buffer pool configurations.

Similar to the earlier configurations in Section \ref{sec: device_character}, 
we use \nvm\ and \ssd\ in one socket and set the CPU affinity mask in SQL Server such that only the CPUs from the same socket are used~\cite{soft-numa-sql}.
We observed that the location of the tempdb where the intermediate results are stored can influence the performance.
Therefore, throughout the experiments, we use the same directory for tempdb which is located in a separate SSD drive not used for persistent data storage.

We study two typical database workloads --- \tpcc\ and \tpch.
For the \tpcc\ experiments, we use OLTPBench \cite{difallah2013oltp, curino2012benchmarking} to issue the queries and configure each run to last 30 minutes.
To minimize the effect of checkpoints, we also set the checkpointing recovery interval in SQL Server to 60 minutes, which effectively disables checkpoints during the \tpcc\ experiments.
For the \tpch\ experiments, we warm up the buffer pool by running each of the 22 queries for multiple times and only consider the runtime for the last execution.

To understand the influence of the dataset size, we use two different scale factors --- SF~100 and SF~500 for \tpch\, and SF~1300 and SF~6500 for \tpcc.
These generate database instances with sizes of approximately 100GB and 500GB respectively.

\subsection{TPC-H benchmark}



\begin{figure}[t]
    \centering
    \includegraphics[width=0.38\textwidth]{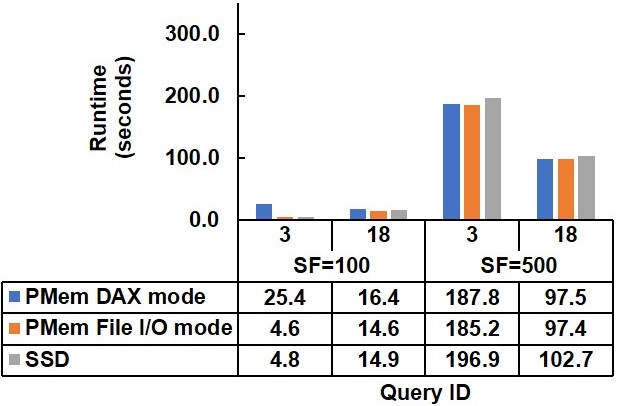}
    \caption{\tpch\ results with \nvm\ as persistent storage}    
    \label{fig:tpch_res1}
\end{figure}

\begin{figure}[t]
    \centering
    \includegraphics[width=0.38\textwidth]{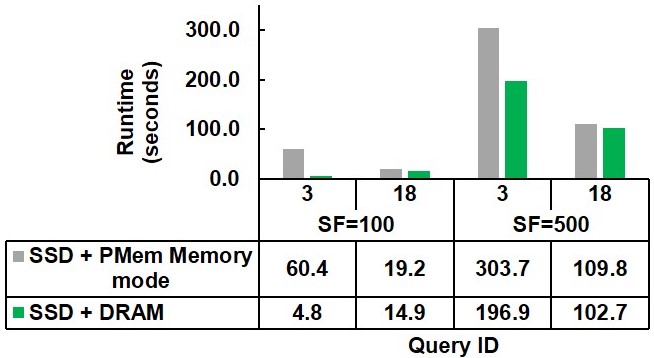}
    \caption{\tpch\ results with \nvm\ as buffer pool}    \label{fig:tpch_res2}
\end{figure}

Due to space constraints, we present a subset of the performance results for \tpch\ queries in Figures~\ref{fig:tpch_res1}--\ref{fig:tpch_res2}. 
We note the following:

\begin{foo}
\textit{With smaller scale factor, i.e. SF=100, \nvm\ \daxio\ fails to outperform \sssd\ and \nvm\ \fileio. As Figure \ref{fig:tpch_res1} shows, when SF=100, the runtime of Query 3 with \nvm\ \daxio\ is \textasciitilde25s, which is \textasciitilde6X slower than \nvm\ \fileio\ and \sssd\  ($<$ 5s for both). This is alleviated with larger SF, i.e. SF = 500, with which \daxio\ is slightly better than \sssd\ (\textasciitilde188s vs. \textasciitilde 197s)}

\textbf{Analysis}: As explained in \cite{hybrid_pmem}, the use of the Hybrid buffer pool with \daxio\ is to reference the pages in \nvm\ instead of caching them in DRAM buffer pool. This implies that with the Hybrid buffer pool enabled, the hot pages in \nvm\ are fetched by CPUs from \nvm\ repetitively, failing to utilize the benefit of the DRAM buffer pool. In contrast, when SF=100, by using {\em \nvm\ \fileio} and {\em \sssd}, large portions of the hot pages are cached in DRAM buffer pool after their first access, leading to smaller overheads in subsequent page accesses and, thus, shorter query execution times. We confirm this phenomenon by monitoring memory usage that reveals that up to 150GB DRAM buffer pool is used for {\em \nvm\ \fileio} and {\em \sssd}, while only \textasciitilde20GB DRAM is used for {\em \nvm\ \daxio}.

In case of larger SFs, only a small portion of the hot pages can be cached in the DRAM buffer pool for {\em \nvm\ \fileio} and {\em \sssd}, which means that the I/O overhead will become more prominent. We observe that Query 3 and Query 18 run slightly faster on \nvm\ due to the smaller I/O overhead compared to {\em \sssd}. Also note that the performance difference between {\em \nvm\ \daxio} and {\em \nvm\ \fileio} is quite small. The reason is that the page size in SQL Server is 8K, for which there is no performance difference under the read workloads between \daxio\ and \fileio\ as observed in Section \ref{sec: micro_bench_storage}.

\textbf{Recommendations} 
This unexpected result indicates the need for further research into good \daxio\ \nvm\ aware page placement policies for the buffer pool in DBMS.
One possible way to improve this is to use similar ideas to the buffer pool extension support in SQL Server \cite{do2011turbocharging} which prioritizes hot pages in DRAM first, and uses \sssd\ as a second chance tier.

\end{foo}

\begin{foo}
\textit{As Figure \ref{fig:tpch_res2} shows, the performance of \sssd\ + \nvm\ \memmode\ is worse than that of \sssd\ + DRAM. For example, with SF=100, the runtime of Query 3 with \sssd\ + \nvm\ \memmode\ is \textasciitilde60s, far longer than the runtime with \sssd\ + DRAM.}
\end{foo}

\textbf{Analysis}: Although \tpch\ is a read-intensive workload, substantial intermediate results are generated during query execution, especially when the data size is large. These are written to the buffer pool and possibly spill to tempdb files if there is not enough memory, resulting in many write operations.
As explained in Section \ref{sec: perf_character_mem}, write operations on \nvm\ \memmode\ can lead to performance loss, thus longer query processing time.

\textbf{Recommendations} 
This result indicates that it is not appropriate to directly replace DRAM with \nvm\ in \memmode\ as the buffer pool in DBMS even for (read intensive) \tpch\ workloads because of the writes to tempdb for intermediate query results.

\subsection{\tpcc\ benchmark}


\begin{figure}[t]
    \centering
    \includegraphics[width=0.33\textwidth, height=0.255\textwidth]{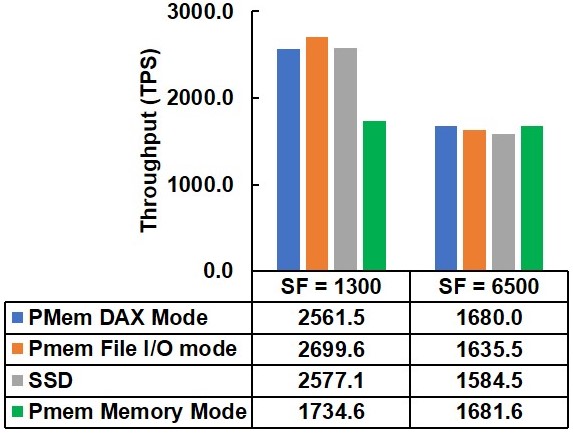}
    \caption{\tpcc\ results with varied scale factors (SF)}    
    \label{fig:tpcc_res}
\end{figure}


We show \tpcc\ performance results in Figure \ref{fig:tpcc_res} and observe that:



\begin{foo}
\textit{There is no significant performance difference between \nvm\ \fileio, \nvm\ \daxio, and \sssd\ under \tpcc\ workloads.}
\end{foo}

\textbf{Analysis}: \tpcc\ is a write-intensive workload. As Figure \ref{fig:bw} in Section \ref{sec: device_character} shows, with the SQL Server page size of 8K, the peak write bandwidth of both {\em \nvm\ \daxio} and {\em \nvm\ \fileio} is only marginally better than that of {\em \sssd}. But when the device I/Os are fully saturated, the write bandwidth of \nvm\ drops compared to the peak write bandwidth due to congestion, which, however, we do not observe in {\em \sssd} (see Figure \ref{fig:bw_thread} in Section \ref{sec: micro_bench_storage}). 

\textbf{Recommendations} The downside of the write operations on \nvm\ has been known by the database community even before the appearance of the real \nvm\ device, which has motivated several related works on limiting the write operations on \nvm\ by redesigning the B-tree index \cite{viglas2012adapting}, join algorithms \cite{viglas2014write}, query optimizer \cite{bausch2012making}, etc, for \nvm-aware DBMSs.
It would be valuable to revisit the feasibility of those early works on the real \nvm\ device.

\section{conclusion and future work}\label{sec: conclusion}
In this paper, we explored some missing device characteristics that are closely related to the database query performance. Our results revealed that some DBMS internal configurations should be changed to take advantage of \nvm\ in the system; (1) A different degree of parallelism, (2) optimal I/O request sizes, and (3) a new page placement policy (to avoid frequent writes on \nvm) should be considered to optimize the use of \nvm\ in DBMSs. 

Our database workload evaluations showed that developers should clearly understand the new device characteristics before introducing software/hardware changes to DBMSs for the best use of \nvm\ in the system. In other words, simple hardware (e.g, replacing DRAM buffer pool with \nvm) or software (e.g, extending DRAM buffer pool with \nvm\ without introducing a new page placement policy) changes will not be properly integrating \nvm\ in DBMSs. 

We revealed some important aspects of using \nvm\ in DBMSs, which can help make better database design choices. However, there are still many open questions on the best use of \nvm\ in the system for many other DBMS internals (e.g., access methods, logging/recovery, etc.).

\newpage
\balance
\bibliographystyle{abbrv}
\bibliography{vldb_sample}

\begin{thebibliography}{10}

\bibitem{FIO315}
{Flexible I/O tester (FIO) rev. 3.15}.
\newblock \url{https://https://fio.readthedocs.io/en/latest/fio_doc.html}.

\bibitem{hybrid_pmem}
{Hybrid buffer pool feature in SQL server}.
\newblock
  \url{https://docs.microsoft.com/en-us/sql/database-engine/configure-windows/hybrid-buffer-pool?view=sql-server-ver15}.

\bibitem{mlc}
{Intel Memory Latency Checker v3.8}.
\newblock
  \url{https://software.intel.com/en-us/articles/intelr-memory-latency-checker}.

\bibitem{pmdk}
Persistent memory development kit.

\bibitem{soft-numa-sql}
{Soft-NUMA (SQL Server)}.
\newblock
  \url{https://docs.microsoft.com/en-us/sql/database-engine/configure-windows/soft-numa-sql-server?view=sql-server-ver15}.

\bibitem{lenovo2019pmem}
{Analyzing the Performance of Intel Optane DC Persistent Memory in App Direct
  Mode in Lenovo ThinkSystem Servers}.
\newblock \url{https://lenovopress.com/lp1083.pdf}, 2019.

\bibitem{lenovo2019pmem2}
{Analyzing the Performance of Intel Optane DC Persistent Memory in Memory Mode
  in Lenovo ThinkSystem Servers}.
\newblock \url{https://lenovopress.com/lp1084.pdf}, 2019.

\bibitem{hpe2019pmem}
{HPE Persistent Memory performance in HPE ProLiant, HPE Synergy, and HPE Apollo
  Gen10 servers with second-generation Intel Xeon Scalable processors}.
\newblock
  \url{https://h20195.www2.hpe.com/v2/getdocument.aspx?docname=a00075993enw},
  2019.

\bibitem{arulraj2019non}
J.~Arulraj and A.~Pavlo.
\newblock Non-volatile memory database management systems.
\newblock {\em Synthesis Lectures on Data Management}, 11(1):1--191, 2019.

\bibitem{arulraj2016write}
J.~Arulraj, M.~Perron, and A.~Pavlo.
\newblock Write-behind logging.
\newblock {\em Proceedings of the VLDB Endowment}, 10(4):337--348, 2016.

\bibitem{bausch2012making}
D.~Bausch, I.~Petrov, and A.~Buchmann.
\newblock Making cost-based query optimization asymmetry-aware.
\newblock In {\em Proceedings of the Eighth International Workshop on Data
  Management on New Hardware}, pages 24--32, 2012.

\bibitem{blelloch2015sorting}
G.~E. Blelloch, J.~T. Fineman, P.~B. Gibbons, Y.~Gu, and J.~Shun.
\newblock Sorting with asymmetric read and write costs.
\newblock In {\em Proceedings of the 27th ACM symposium on Parallelism in
  Algorithms and Architectures}, pages 1--12. ACM, 2015.

\bibitem{chen2015persistent}
S.~Chen and Q.~Jin.
\newblock Persistent b+-trees in non-volatile main memory.
\newblock {\em Proceedings of the VLDB Endowment}, 8(7):786--797, 2015.

\bibitem{curino2012benchmarking}
C.~A. Curino, D.~E. Difallah, A.~Pavlo, and P.~Cudre-Mauroux.
\newblock Benchmarking oltp/web databases in the cloud: The oltp-bench
  framework.
\newblock In {\em Proceedings of the fourth international workshop on Cloud
  data management}, pages 17--20, 2012.

\bibitem{difallah2013oltp}
D.~E. Difallah, A.~Pavlo, C.~Curino, and P.~Cudre-Mauroux.
\newblock Oltp-bench: An extensible testbed for benchmarking relational
  databases.
\newblock {\em Proceedings of the VLDB Endowment}, 7(4):277--288, 2013.

\bibitem{do2011turbocharging}
J.~Do, D.~Zhang, J.~M. Patel, D.~J. DeWitt, J.~F. Naughton, and A.~Halverson.
\newblock Turbocharging dbms buffer pool using ssds.
\newblock In {\em Proceedings of the 2011 ACM SIGMOD International Conference
  on Management of data}, pages 1113--1124, 2011.

\bibitem{imamura2020analysis}
S.~Imamura and E.~Yoshida.
\newblock The analysis of inter-process interference on a hybrid memory system.
\newblock In {\em Proceedings of the International Conference on High
  Performance Computing in Asia-Pacific Region Workshops}, pages 1--4, 2020.

\bibitem{izraelevitz2019basic}
J.~Izraelevitz, J.~Yang, L.~Zhang, J.~Kim, X.~Liu, A.~Memaripour, Y.~J. Soh,
  Z.~Wang, Y.~Xu, S.~R. Dulloor, et~al.
\newblock Basic performance measurements of the intel optane dc persistent
  memory module.
\newblock {\em arXiv preprint arXiv:1903.05714}, 2019.

\bibitem{kadekodi2019splitfs}
R.~Kadekodi, S.~K. Lee, S.~Kashyap, T.~Kim, A.~Kolli, and V.~Chidambaram.
\newblock Splitfs: reducing software overhead in file systems for persistent
  memory.
\newblock In {\em Proceedings of the 27th ACM Symposium on Operating Systems
  Principles}, pages 494--508, 2019.

\bibitem{lersch2019evaluating}
L.~Lersch, X.~Hao, I.~Oukid, T.~Wang, and T.~Willhalm.
\newblock Evaluating persistent memory range indexes.
\newblock {\em Proceedings of the VLDB Endowment}, 13(4):574--587, 2019.

\bibitem{lersch2019persistent}
L.~Lersch, W.~Lehner, and I.~Oukid.
\newblock Persistent buffer management with optimistic consistency.
\newblock In {\em Proceedings of the 15th International Workshop on Data
  Management on New Hardware}, pages 1--3, 2019.

\bibitem{mironov2019performance}
V.~Mironov, I.~Chernykh, I.~Kulikov, A.~Moskovsky, E.~Epifanovsky, and
  A.~Kudryavtsev.
\newblock Performance evaluation of the intel optane dc memory with scientific
  benchmarks.
\newblock In {\em 2019 IEEE/ACM Workshop on Memory Centric High Performance
  Computing (MCHPC)}, pages 1--6. IEEE, 2019.

\bibitem{peng2019system}
I.~B. Peng, M.~B. Gokhale, and E.~W. Green.
\newblock System evaluation of the intel optane byte-addressable nvm.
\newblock In {\em Proceedings of the International Symposium on Memory
  Systems}, pages 304--315, 2019.

\bibitem{psaropoulos2019bridging}
G.~Psaropoulos, I.~Oukid, T.~Legler, N.~May, and A.~Ailamaki.
\newblock Bridging the latency gap between nvm and dram for latency-bound
  operations.
\newblock In {\em Proceedings of the 15th International Workshop on Data
  Management on New Hardware}, pages 1--8, 2019.

\bibitem{van2018managing}
A.~van Renen, V.~Leis, A.~Kemper, T.~Neumann, T.~Hashida, K.~Oe, Y.~Doi,
  L.~Harada, and M.~Sato.
\newblock Managing non-volatile memory in database systems.
\newblock In {\em Proceedings of the 2018 International Conference on
  Management of Data}, pages 1541--1555. ACM, 2018.

\bibitem{van2019persistent}
A.~van Renen, L.~Vogel, V.~Leis, T.~Neumann, and A.~Kemper.
\newblock Persistent memory i/o primitives.
\newblock In {\em Proceedings of the 15th International Workshop on Data
  Management on New Hardware}, pages 1--7, 2019.

\bibitem{viglas2012adapting}
S.~D. Viglas.
\newblock Adapting the b+-tree for asymmetric i/o.
\newblock In {\em East European Conference on Advances in Databases and
  Information Systems}, pages 399--412. Springer, 2012.

\bibitem{viglas2014write}
S.~D. Viglas.
\newblock Write-limited sorts and joins for persistent memory.
\newblock {\em Proceedings of the VLDB Endowment}, 7(5):413--424, 2014.

\bibitem{weiland2019early}
M.~Weiland, H.~Brunst, T.~Quintino, N.~Johnson, O.~Iffrig, S.~Smart, C.~Herold,
  A.~Bonanni, A.~Jackson, and M.~Parsons.
\newblock An early evaluation of intel's optane dc persistent memory module and
  its impact on high-performance scientific applications.
\newblock In {\em Proceedings of the International Conference for High
  Performance Computing, Networking, Storage and Analysis}, pages 1--19, 2019.

\bibitem{yang2019empirical}
J.~Yang, J.~Kim, M.~Hoseinzadeh, J.~Izraelevitz, and S.~Swanson.
\newblock An empirical guide to the behavior and use of scalable persistent
  memory.
\newblock {\em arXiv preprint arXiv:1908.03583}, 2019.

\bibitem{zarubin2019integer}
M.~Zarubin, P.~Damme, T.~Kissinger, D.~Habich, W.~Lehner, and T.~Willhalm.
\newblock Integer compression in nvram-centric data stores: Comparative
  experimental analysis to dram.
\newblock In {\em Proceedings of the 15th International Workshop on Data
  Management on New Hardware}, pages 1--11, 2019.

\bibitem{zarubin2019efficient}
M.~Zarubin, T.~Kissinger, D.~Habich, T.~Willhalm, and W.~Lehner.
\newblock Efficient compute node-local replication mechanisms for nvram-centric
  data structures.
\newblock {\em The VLDB Journal}, pages 1--21, 2019.

\end{thebibliography}

\end{document}